\documentclass[prl,showpacs,aps,superscriptaddress,twocolumn]{revtex4}

\usepackage{amsmath, amssymb}
\usepackage{graphicx}
\usepackage{graphics}
\usepackage{dsfont}
\newcommand{\void}[1]{}

\begin{document}
\title{On the Dynamical detection of Majorana fermions in current-biased nanowires}

\author{Fernando \surname{Dom\'inguez}}
\affiliation{Instituto de Ciencia de Materiales, CSIC, Cantoblanco, E-28049 Madrid, Spain}
\author{Fabian  \surname{Hassler}}
\affiliation{Institute for Quantum Information, RWTH Aachen University, 52056 Aachen, Germany}
\author{Gloria \surname{Platero}}
\affiliation{Instituto de Ciencia de Materiales, CSIC, Cantoblanco, E-28049 Madrid, Spain}

\date{\today}


\begin{abstract}
We analyze the current-biased Shapiro
experiment in a Josephson junction formed by two one-dimensional
nanowires featuring Majorana fermions.  Ideally, these junctions are
predicted to have an unconventional $4\pi$-periodic Josephson effect and
thus only Shapiro steps at even multiples of the driving frequency. Taking
additionally into account overlap between the Majorana fermions, due to the
finite length of the wire, renders the Josephson junction conventional for
any dc-experiments. We show that probing the current-phase relation in a
current biased setup dynamically decouples the Majorana fermions. We find
that besides the even integer Shapiro steps there are additional steps at
odd and fractional values. However, different from the voltage biased case,
the even steps dominate for a wide range of parameters even in the case of
multiple modes thus giving a clear experimental signature of the presence
of Majorana fermions.
\end{abstract}
\maketitle

Majorana Fermions (MFs) have recently been predicted to occur in a
multitude of different condensed-matter systems \cite{Volovik1989a,
Moore1991a, Read2000a, Kitaev2003a, Fu2008a, Lutchyn2010a, Oreg2010a,
Choy2011a}. The interest in MFs stems from the non-Abelian quantum
statistics which forms the basis of topological quantum computation
\cite{Moore1991a, Ivanov2001a,Kitaev2003a, Nayak2008a}. 
Majorana fermions naturally occur in half-vortices of chiral
\emph{p}-wave superconductors. Although this type of superconductivity
has not been found, it was realized recently that
\emph{s}-wave superconductor together with strong spin-orbit and an applied
magnetic field may emulate a \emph{p}-wave superconductor \cite{Fu2008a,
Lutchyn2010a, Oreg2010a, Linder2010a}. During the last months three 
different experiments \cite{Mourik2012a, Rodrigo2012a, Rokhinson2012a} 
appeared in the literature which may provide the first experimental 
evidence of MFs.

Signatures of Majorana Fermions appear in the electrical
\cite{Bolech2007a,Law2009a, Wimmer2011a} and thermal conductance
\cite{Wimmer2010a,Akhmerov2011a}, shot-noise \cite{Akhmerov2011a},
Andreev-reflection \cite{Law2009a,Flensberg2010a} and the non-local
tunneling \cite{Nilsson2008a,Law2009a,Fu2010a,Benjamin2010a}. 
In this Letter we will focus on the measurement of the 
fractional Josephson effect, given when we put together two superconductors
featuring MFs \cite{Kitaev2001a, Kwon2003a, Fu2009a, Tanaka2009a, 
Lutchyn2010a,Oreg2010a,Ioselevich2011a,Law2011a,Heck2011a}.
Physically, this effect is produced by the fact that in the 
presence of a Majorana bound mode, the supercurrent carries 
single electrons instead of the usual Cooper pairs. 
Thus, this fractional Cooper pairs affect the 
supercurrent by turning it from $\sin(\varphi)$ to $\sin(\varphi/2)$.

In Josephson junctions, Shapiro step
experiments allow for the deduction of the periodicity of the current-phase
relation of the junction \cite{Tinkham1996a,Shapiro1963a}. Very recently,
Shapiro-steps have been analyzed for voltage-biased Majorana wires
\cite{Jiang2011a, San-Jose2011a, Pikulin2011b}. However, the more 
experimentally realistic current-biased experiment \cite{Tinkham1996a} 
remains unexplored.

In one-dimensional (1D) Majorana wire, MFs will appear at the end points
\cite{Kitaev2001a}. In an ideal situation, the ends are infinitely apart
from each other avoiding their recombination. In turn, when the wire is
finite, the overlap, although very small, is different from zero, thus MF
pair recombines and the special properties that the MF confer to the system
are lost immediately \cite{Pikulin2011a}.
Physically, one can circumvent this problem
using a Josephson junction where the gauge invariant phase is tuned non-adiabatically.
In this way, transitions between the recombined 
fermions induce a dynamical decoupling into Majorana fermions.

\begin{figure}[tb]
\begin{center}
\includegraphics[width=2.5in,clip]{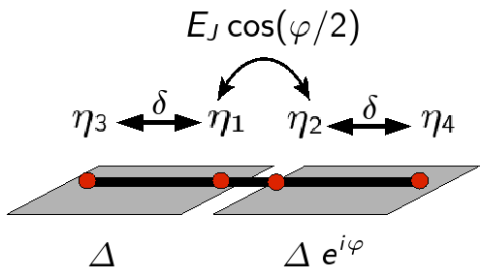}
\end{center}
\caption {\label{Fig.schema} Josephson junction with a nanowire on top. 
Red spots (color online) represent Majorana fermions. Double arrows 
represent the overlap between the Majorana fermions.}
\end{figure}

In this work we analyze theoretically the current biased Shapiro experiment
\cite{Shapiro1963a} in a finite 1D Josephson junction where the MFs are
recombined (see Fig.~\ref{Fig.schema}). In the presented setup, the current bias the 
gauge invariant phase, inducing dynamical decoupling of the MFs. 
Meanwhile, it induces a voltage 
difference that can be measured, presenting the pattern of the periodicity of 
the junction. 
We have calculated the induced voltage by means of the 
Resistively Shunted Junction (RSJ) model \cite{Tinkham1996a}. 
In addition, we include extra Andreev modes carrying a 2$\pi$
periodic current and quasiparticle poisoning (QP). 

\begin{figure}[tb]
\begin{center}
\includegraphics{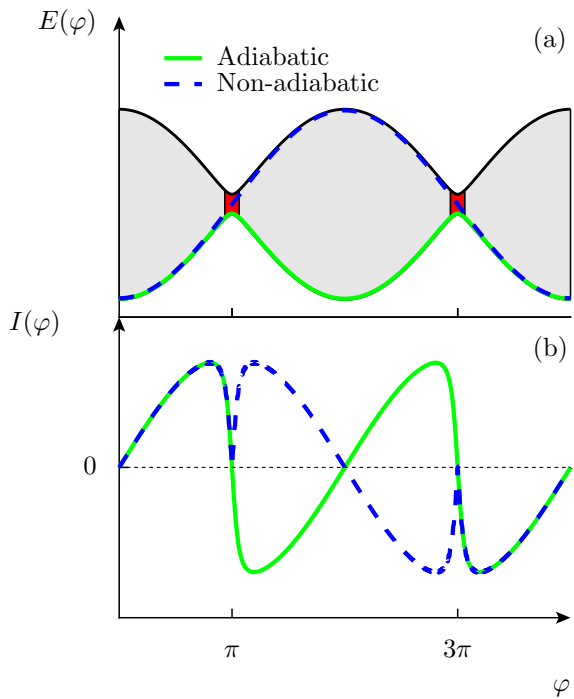}
\end{center}
\caption {\label{Fig.energy} (Color on line) Eigenspectrum (a) and Andreev 
current (b) vs $\varphi$, after zero (green solid) and two (blue dashed) LZT occurred. 
In panel (a) light and dark gray areas (gray and red) correspond to 
the adiabatic and non-adiabatic evolution respectively. We have represented 
two different limits of dynamics; the adiabatic limit (green solid line), 
presenting 2$\pi$ periodicity and the non-adiabatic limit (blue dashed line) 
presenting 4$\pi$ periodicity.}
\end{figure}

In contrast to the
infinite length case, where only even Shapiro steps appear, the obtained
results show small contribution steps at odd and fractional multiples of the 
ac frequency, coming from the
new features of the dynamical current (see Fig.~\ref{Fig.energy}(b) below). 
Nevertheless, all these contributions are of the order of the overlap between 
in-phase MFs and 
negligible compared to the height 
of the even Shapiro steps.
Remarkably, we have found a regime where the effect of considering
a dominant contribution of extra 2$\pi$ periodic Andreev modes, 
does not modify the spectrum of even Shapiro steps, 
providing a robust measurement of the 4$\pi$ periodicity. 
In addition, we have seen that typical time scales ($\mu$s) 
\cite{Sun2011a} of QP produce a negligible effect on the dynamics of the system.

Ideally, a generic 1D Josephson junction in the presence of Majorana
fermions can be described by the MFs placed at the junction,
yielding a Hamiltonian
\begin{align}
H_0={\it i}E_J \cos\left(\varphi/2\right)\eta_1\eta_2,
\label{eq:1dideal}
\end{align}
where $E_J$ is the Josephson energy of the junction, $\varphi$ is the gauge
invariant phase difference and the operators $\eta_i$ are Hermitian
$\eta_i=\eta_i^\dagger$ and they fulfill the anticommutator relation 
$\eta_i \eta_j+\eta_j \eta_i=2\delta_{i,j}$. 
Due to the presence of MF the periodicity of the spectrum is 4$\pi$.
In finite systems, in-phase MF may recombine into usual fermions through
 the overlap of their wave functions \cite{Kitaev2001a, Pikulin2011a}. 
 In order to account with this phenomenon an extra term should be added,
  so that the total Hamiltonian becomes
\begin{align}
H={\it i}E_J \cos\left(\varphi/2\right)\eta_1\eta_2+
{\it i}\delta\left(\eta_4\eta_2+\eta_1\eta_3\right),
\label{eq:H1}
\end{align}
where we have introduced a parameter $\delta$ to account
for overlap between the in-phase MF which decreases exponentially with
increasing distance between the Majorana modes
(see Fig.~\ref{Fig.schema}). Considering that the in-phase MFs are
far away compared with those on the junction we will use $\delta\ll E_J$.
Diagonalizing the Hamiltonian yields the 2$\pi$-periodic energy spectrum
(see Fig.~\ref{Fig.energy}(a))
\begin{align}
E(\varphi)=\pm\sqrt{4\delta^2+E_J^2\cos^2(\varphi/2)}.
\label{eq:eigenspectrum}
\end{align}

Non-adiabatic changes of the phase leads to transitions between the
two eigenstates. Since $E_J\gg \delta$, the transition probability
is non-vanishing only at the anticrossings of the eigenspectrum,
that is, for $\varphi=(2n+1)\pi$, where $n$ is an integer (see red areas in
Fig.~\ref{Fig.energy}(a)). Thus, as long as non-adiabatic transitions
occur, the overlap between MF is effectively
canceled. As a consequence, the $4\pi$ periodicity in the eigenspectrum,
and also in the supercurrent ($I\propto \partial_\varphi E_{\pm}$), is
recovered. As we will see below the new shape of the current does lead to 
the expected even steps and also to additional contributions of the order of 
$\delta$ at odd and fractional multiples of the 
ac frequency \cite{supplementary}.

In order to calculate the transition probability we consider
the semiclassical approximation, and we make use of the 
fact that the velocity at the anticrossings is linear, therefore, 
transitions between states can be obtained by means of 
the Landau-Zener probability
\begin{align}
P_{LZ}=\exp\left(-2\pi\frac{4\delta^2}{E_J \hbar\dot{\varphi}}\right).
\end{align}
It is important to remark that in the experiment we are analyzing, the phase 
$\varphi$ is biased by a noisy voltage coming from fixing an external current. 
These voltage fluctuations are translated to phase fluctuations by the fact
that $\dot{\varphi}\propto V$, and thus, dephasing enters into play. We have 
estimated  that the dephasing time $t_D$ 
is much shorter than the time needed to change the phase by 
$\varphi\rightarrow \varphi +2\pi$. 
Therefore, we assume that
interference effects can be neglected, and Landau-Zener
transitions (LZT) can be considered individually. 
Coherences between LZTs have been recently analyzed 
phenomenologically \cite{Pikulin2011b}, and 
in more detail \cite{San-Jose2011a} for the case of
a voltage biased junction, where also 
additional Andreev levels, QP and inelastic
transitions have been considered. 
However, we would like to stress that the current biased
setup analyzed in this letter for the first time has two advantages:
contrary to the voltage biased case, it (1) shows a robust signal of small
odd integer Shapiro steps even if the case of a multimode wire and (2)
the observation is not masked by interference effects as those are absent
in our case.

Once we have analyzed the dynamical transitions of the junction, we 
are ready to include their dynamical effects on the current. To this 
aim we introduce the function $I_M(\varphi)$ in the supercurrent 
\begin{align}
I(\varphi)=I_M(\varphi)\frac{2}{E_J}\frac{\partial}{\partial \varphi}E (\varphi).
\label{eq.dynsc}
\end{align}
The function, $I_M(\varphi)$, can take the constant values $\pm I_M$, 
where $I_M$ is the maximum value of the supercurrent, which is of the 
order of nA. During the adiabatic period, the function $I_M(\varphi)$ 
remains constant and whenever there is a LZT, $I_M(\varphi)$ changes 
its sign. To understand the change of the sign we can compare in 
Fig.~\ref{Fig.energy}(b) the adiabatic and non-adiabatic passage 
through the anticrossings (solid and dashed respectively). After each 
anticrossing the curve coming from a LZT acquires a negative sign respect 
to the adiabatic passage. Thus, we describe the dynamical effects on the 
current produced by the LZT by changing the sign of $I_M$. 

We study the Shapiro experiment by means of the resistive shunted junction model 
(RSJ) in the overdamped limit \cite{Tinkham1996a, Russer1972a}. The induced voltage 
on the junction can be calculated by solving the differential equation
\begin{align}
I_0+I_1\sin(\omega_{ac} t)=I(\varphi(t))+\frac{\hbar}{2e R}\dot{\varphi}(t).
\label{eq:RSJ}
\end{align}
This equation is obtained from Kirchoff's law where an external DC $I_0$ and
AC $I_1 \sin(\omega_{ac} t)$ currents are applied to the junction. The
outgoing current is modeled by a parallel circuit whose components are,
$I(\varphi(t))$, given by Eq.~(\ref{eq.dynsc}), and a resistive current
$(\hbar/2e R)\dot{\varphi}$ originating
from the existence of quasiparticles. 
The solution of the differential equation (\ref{eq:RSJ}), allows to obtain the
induced voltage $V = \hbar\dot{\varphi}/2e$. This equation is solved 
dynamically since $I_M(\varphi)$ changes its sign depending on 
whether the LZT occurs or not. In order to include such a dynamical effect 
we compute $P_{LZ}$, on the anticrossings. Namely, when $\varphi(t)=(2n+1)\pi$ 
we evaluate the phase velocity $\dot{\varphi}$ and compare the resulting 
$P_{LZ}$ with a random number to determine whether a LZT occurs or not.

In order to perform the calculations closer to the real experiment
we have included two additional phenomena: the presence of extra Andreev
modes and QP. Majorana fermions may occur in
multimode nanowires as long as there is an odd-number of bands occupied
\cite{Wimmer2010a,Potter2010a}. The extra modes contribute with the conventional
2$\pi$-periodic supercurrent, i.e.,~by means of adding $I_c\sin(\varphi)$ to
$I(\varphi)$ in Eq.~(\ref{eq.dynsc}). 
The value $I_c$ is a constant parameter
whose contribution can be much greater than $I_M$, due to the possibility
that several modes are occupied. 
It is worth to mention that we have not considered dynamical effects between the 
Majorana and Andreev modes due to the fact that the energy difference between them
is high enough compared to the phase velocity. This can be assured whenever the 
transparency of the junction fulfils $T\ll 1$ \cite{Fu2009a,San-Jose2011a}. 
The second effect is the QP, which accounts
for the effect of quasiparticles tunneling from the contacts. 
The change
in the number of quasiparticles provokes a transition from one of the
eigenstates to the other one \cite{Fu2008a}. These transitions affect the
current periodicity, and in principle they should appear reflected on the
Shapiro steps positions.

\begin{figure}[tb]
\begin{center}
\includegraphics[width=3.0in,clip]{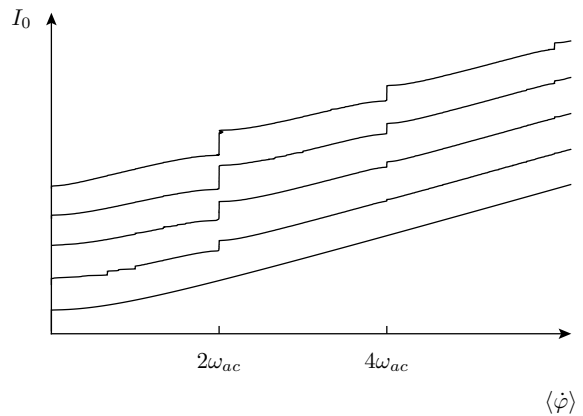}
\end{center}
\caption {\label{Fig.shapiro} 
Current-Voltage curves for $I_M=1\,$nA, $E_J/\delta=500$,
$R=3 \,\text{k}\Omega$, $\omega_{ac}=10^{10}$Hz and $I_c=0$.
$I_1$ takes values from 0 to 4$\,$nA from bottom 
to top with a step of 1$\,$nA.  
For the sake of clarity all the curves are shifted a constant amount.
}
\end{figure}%

\begin{figure}[tb]
\begin{center}
\includegraphics[width=3.0in,clip]{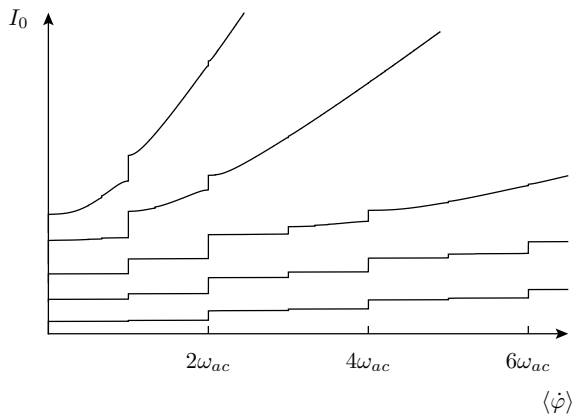}
\end{center}
\caption {\label{Fig.shapiro2} Current-Voltage curves for 
$I_M=1\,$nA, $E_J/\delta=500$,
$R=3 \,\text{k}\Omega$, $I_1=3\,$nA and $I_c=10\, I_M$.
The ac frequency $\omega_{ac}$ takes the values: $10^{11}$, $5\times 10^{10}$, 
$2\times 10^{10}$, $10^{10}$ and $5 \times 10^{9}\,$Hz from top to bottom.
We observe a progressive quenching of the odd steps as we decrease $\omega_{ac}$
up to $10^{10}\,$Hz.
For the sake of clarity all the curves are shifted down a constant amount.}
\end{figure}

The main results of our calculations are shown in Figs.~\ref{Fig.shapiro} 
and~\ref{Fig.shapiro2}. 
There, we have represented current-voltage curves for $I_c=0$ and $I_c=10\,I_M$ 
respectively, using $I_M=1\,$nA and 
R=$3\,\text{k}\Omega$, altogether lead to measure voltages of the order 
of $V\approx 10\,\mu$V. 
We have set $E_J/\delta=500$, making that $P_{LZ}$ 
is very close to one at any value of $I_0$. Therefore, the supercurrent  
presents a 4$\pi$ periodicity most of the time, whose form is given by
the dashed curve of Fig.~\ref{Fig.energy}(b).

In figure~\ref{Fig.shapiro} we have plotted the $I/V$ curves for 
$\omega_{ac}=10^{10}\,$Hz and $I_1=0$ up to $I_1=4$nA
with an increment of 1$\,$nA. We can observe that the height of even Shapiro steps
dominate the Shapiro spectrum. Thus, revealing clearly the 
resulting 4$\pi$ periodicity of the junction.
Moreover, there are some contributions of $\delta$ order placed
at odd and fractional multiples of $\omega_{ac}$, coming from the new form of 
the supercurrent acquired by the LZT.
It is worth to remind that in the ideal case, 
i.e.~$I(\varphi)=I_M\sin(\varphi/2)$ such steps do not appear. 

In figure~\ref{Fig.shapiro2} we show the $I/V$ curves in the presence of a 
large 2$\pi$ contribution, $I_c=10\, I_M$. Different curves correspond to
different values of $\omega_{ac}$, in decreasing 
order from top to bottom. In this situation, one would expect to obtain all integer
steps coming from the dominant 2$\pi$ character of the supercurrent. 
In turn, one observes a progressive reduction of the odd steps as we decrease 
$\omega_{ac}$ up to $2e R I_M/\hbar$, while even steps hold. 
We have demonstrated that the reduction of odd steps is caused by the presence 
of a Majorana mode and can, in principle, be found even in the case where $I_c\gg I_M$.
It is important to remark that this behavior is completely different 
to the voltage biased experiment, 
where the 2$\pi$ contribution gives rise to steps at even and odd multiples, 
with heights proportional to $I_c$ \cite{Jiang2011a}. 
Thus, when $I_c\gg I_M$ the detection of the Majorana mode from the Shapiro spectrum 
will in general complicated.

The observed behavior stands on 
the non-linear character of the RSJ model versus the linear one 
of the voltage biased experiment.
In order to understand this it is 
necessary first to revisit the undriven case, i.e.~$I_1=0$, where both, 
2$\pi$ and 4$\pi$ contributions are present.
In such situation, we show \cite{supplementary} 
that due to the non-linear character of the RSJ equation, the presence
of the $4\pi$ contribution imposes a strong $4\pi$ character to the
junction for a range of voltages of the order of $I_M R$, even in the case
when $I_c \gg I_M$. Ac currents with frequencies up to $2eRI_M /\hbar=
10^{10}\,$Hz, correspond to this range of voltages and thus are expected
to show predominately even Shapiro step.

We can extract additional information from the current-voltage curves. 
It can be demonstrated that the height of the 0-step at $I_1=0$, is approximately
equal to $I_c+I_M/\sqrt{2}$ for $I_c\gg I_M$. Then, tuning the gate voltage 
one could, in principle, fill a single extra mode and measure the resulting 
contribution to the current. Then, one would be able to determine $I_M$.

\begin{figure}[tb]
\begin{center}
\includegraphics{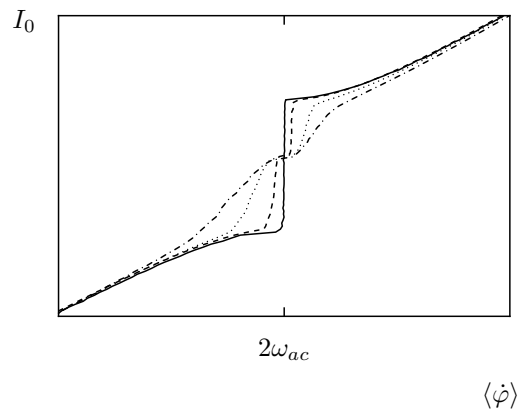}
\end{center}
\caption {\label{Fig.zoom} Current voltage curves for different values of
$E_J/\delta$=500 (solid), 100 (dashed), 50 (dotted) and
30 (dotted-dashed), $I_1=1\,$nA, the rest of the parameters
are taken from those used in Fig.~\ref{Fig.shapiro}. 
We show the splitting of a Shapiro step by the increment of the overlap $\delta$}
\end{figure}

Until now the calculations we have shown were performed using the value
$E_J/\delta=500$, which leads to $P_{LZ}\simeq1$ for the whole spectrum of $I_0$.
Increasing the overlap $\delta$ reduces $P_{LZ}$ making that non-LZT may
occur, provoking a departure from the studied $4\pi$ periodicity of the current.
The main changes are produced in the positions of the steps, where we observe
that the Shapiro step splits in two (see Fig.~\ref{Fig.zoom}).
In order to shed some light on the numerical results we can average out the 
position of the non-LZT events and approximate $I_M(\varphi)$ by means of a 
Fourier series \cite{supplementary}. 
The resulting 
current-voltage curve behaves rather similar to the numerical results obtained 
by means of the stochastic model presented here. 
Comparing both methods we extract 
that the splitting is of the order of $(1-P_{LZ})/2$. We observe that the 
Fourier approximation fails whenever we decrease $E_J/\delta$ up to 30.

Quasiparticle poisoning relaxation time is typically of the order 
of $\mu$s \cite{Sun2011a}. For the ac driving
frequencies considered in our work, $\omega_{ac}\approx 10^{10}$Hz, QP does
not affect  our results significantly.

In conclusion, we have studied the current-biased Shapiro experiment
in a finite 1D Josephson junction. We have seen that the effects coming
from the finiteness of the 1D system are dynamically decoupled driving
the phase by means of an external current. For that purpose,
we have analyzed the periodicity of the junction by solving an equation of
motion coming from the resistive shunted junction model. 
We have calculated two different cases, when there are no extra 2$\pi$ modes, 
i.e.~$I_c=0$, we have obtained that we can always determine the presence of
the Majorana mode due to the appearance of steps at even multiples of the ac 
frequency $\omega_{ac}$ (see Fig.~\ref{Fig.shapiro}). 
In turn, when $I_c\gg I_M$, we have found that thanks to the 
non-linear effects coming from the dynamics of RSJ, the junction behaves
4$\pi$ periodically for a range of voltages of the order $I_M R$. 
We have found that it is possible to match this range, 
and therefore its 4$\pi$ behavior,
by using ac frequencies of the order of 
$\omega_{ac}\lesssim 2e R I_M/\hbar=10^{10}$Hz (see Fig.~\ref{Fig.shapiro2}).
The resulting Shapiro steps are thus placed at even multiples of the frequency
$\omega_{ac}$.
In addition,
we have seen that the effects of QP on the current
are negligible at the typical estimated timescales. From our results, we
believe that performing the current-biased Shapiro experiment is a strong
tool to prove the existence of Majorana fermions in finite nanowires.

\section{ACKNOWLEDGMENTS}
We acknowledge R. Hussein for his invaluable help in improving the numerical codes.
We acknowledge L. P. Kouwenhoven and S. Frolov for enlightening discussions.
One of the authors (F.D) acknowledges  the hospitality of the  Kavli Institute of 
Nanoscience Delft. We also acknowledge financial support from Grant No. MAT2011-24331
and from ITN, Grant No. 234970 (EU). 
FH is grateful for support from the Alexander von Humboldt foundation.


\clearpage
\appendix
\widetext

\section{Supplementary Material to `On the dynamical detection 
of Majorana fermions in finite one-dimensional systems'}

\section{Effective Hamiltonian and eigenvalues of a finite Josephson junction}

We start from a more general Hamiltonian than that given in Eq.~(2)
of the manuscript. This kind of generic Hamiltonian has been
already used previously, to describe a wire in the presence of
Majorana fermions overlapping between each other. See for example
Refs.~\onlinecite{Kitaev2001a} or~\onlinecite{Leijnse2010a}.  We then write,
\begin{align}
H={\it i}\epsilon \eta_1\eta_2+
{\it i}\delta_R \eta_4\eta_2+{\it i}\delta_L\eta_1\eta_3.
\label{eq:H1}
\end{align}
In contrast to Eq.~(2), we assume that the overlap $\delta_{L,R}$ between
the different sides of the junction can be in general different, which
is something that is more similar to a real system. Besides, we simplify
the notation by using $\epsilon=E_J \cos\left(\varphi/2\right)$.  In this
situation it is useful to switch to a representation where two Majorana
fermions are combined to form one ordinary fermion. Thus, performing
the substitutions $\eta_1={\it i}(l^\dagger-l)$, $\eta_2=r^\dagger+r$,
$\eta_3=l^\dagger+l$, and $\eta_4={\it i}(r^\dagger-r)$, we obtain
\begin{align}
H=2\delta_R \left(r^\dagger r-\frac{1}{2}\right)+ 2\delta_L \left(l^\dagger l-\frac{1}{2}\right)+\epsilon \left(-l^\dagger r^\dagger - l^\dagger r+l r^\dagger+lr\right),
\end{align}
where $l^\dagger l$ and $r^\dagger r=0,~1$ counts the occupation of
the corresponding state. We rewrite the former Hamiltonian in the base
of occupation $|n_L,n_R\rangle$, that is $|0,0\rangle$, $|0,1\rangle$,
$|1,0\rangle$ and $|1,1\rangle$, yielding the matrix
\begin{align}
H=
\begin{pmatrix}
                 -\delta_L-\delta_R  & 0 &  0 & -\epsilon \\
                 0  & \delta_R-\delta_L  &  -\epsilon & 0 \\
		0  & -\epsilon  & \delta_L-\delta_R & 0 \\
		-\epsilon  & 0  &  0 & \delta_L+\delta_R 
\end{pmatrix}.
\end{align}
We diagonalize the resulting Hamiltonian and obtain 4 eigenvalues
\begin{align}
&E^e_\pm=\pm\sqrt{\epsilon^2+ \left(\delta_L+\delta_R \right)^2}\text{ and}\\
&E^o_\pm=\pm\sqrt{\epsilon^2+ \left(\delta_L-\delta_R \right)^2}.
\end{align}
where the super index denotes even (e) and odd (o) fermion parity
respectively.  Assuming that the parity is conserved we can choose one of the
eigenvalues and make Landau-Zener transitions between the $\pm$ eigenvalues.
In fact, we are only interested in the generic features which an avoided
crossing has on the Shapiro steps. Thus we replace either $(\delta_L \pm
\delta_R)^2$ by a characteristic/generic value $2\delta$ which directly
leads to Eq.~(3). We have included a footnote in the manuscript making this
explicit. Note furthermore that the signs of $\delta_{L/R}$ are random,
i.e., determined by microscopic details. Thus, it is not true that the odd
sector has a smaller gap than the even sector as one might naively expect
when looking at the equations above.

\section{Dephasing produced biasing the phase by means of a noisy voltage}

In the main text we have used the approximation of neglecting interference 
effects coming from having two or more non-adiabatic 
transitions \cite{Shevchenko2010a}. In this appendix we justify this approximation 
and estimate the dephasing rate produced by biasing the phase difference by 
means of a fluctuating voltage. For the sake of simplicity we restrict the 
analysis to the infinite length 1D Josephson junction which in the pseudo-spin 
basis is given by
\begin{align}
H=\frac{E_J}{2}\cos(\varphi/2)\sigma_z,
\end{align}
where $\sigma_z$ denotes the {\it z}-Pauli matrix.
As we have explained, the effect of fixing the current 
produces thermal fluctuations in the voltage, therefore 
fluctuations on the phase difference arise around some fixed 
value $\varphi_0$, namely $\varphi(t)=\varphi_0+\delta\varphi(t)$. 
Then, we assume that these fluctuations are small compared with 
$\varphi_0$, so we can rewrite the Hamiltonian approximately
\begin{align}
H\approx \frac{E_J}{2}\left(\cos(\varphi_0/2)\sigma_z-
\delta\varphi(t)\sin(\varphi_0)\sigma_z\right).
\end{align}
Thus, the energy difference between the states fluctuates in time. 
In order to see the effects of these fluctuations one can take a 
coherent superposition such as 
\begin{align}
\frac{1}{\sqrt{2}}\left(|+\rangle + |-\rangle \right),
\end{align}
where $|\pm\rangle$ are the eigenstates of $\sigma_z$. 
The state at time t is given by
\begin{align}
\frac{1}{\sqrt{2}}\left(e^{-{\it i}(\Delta t + \phi(t))}|
+\rangle + e^{{\it i}(\Delta t + \phi(t))}|-\rangle\right).
\end{align}
Where we have used $\Delta$ to denote the constant energy difference between 
the states $|\pm\rangle$. 
The phase $\phi(t)$ accounts the time integral over the fluctuating 
component of the energy difference
\begin{align}
\phi(t)= \frac{E_J\sin(\varphi_0)}{\hbar}\int_0^t d\tau \delta\varphi(\tau).
\end{align}
These fluctuations are seen in the expectation value
\begin{align}
\langle\sigma_x\rangle(t)=\cos(\Delta t + \phi(t)).
\end{align}
Taking the average over the fluctuating phase we observe a 
decay of the oscillations
\begin{align}
\langle\langle\sigma_x\rangle(t)\rangle = \cos(\Delta t)
e^{-\frac{1}{2}\langle \phi(t)^2\rangle}.
\end{align}
where 
\begin{align}
\langle \phi(t)^2\rangle=\int_0^{t}dt_1\int_0^{t}dt_2\langle 
\delta\varphi(t_1)\delta\varphi(t_2)\rangle,
\end{align}
is the time integral of the autocorrelation function. Obtaining 
this integral allows us to determine lifetime of the coherences. 
In order to calculate it we have first to relate phase and 
voltage autocorrelation functions through the Josephson formula
\begin{align}
\frac{d}{dt}\varphi(t)=\frac{2e}{\hbar}V.
\end{align}
So that we can express the phase correlation function in terms 
of the voltage correlation function can be written as
\begin{align}
\langle \phi(t)^2\rangle=\left(\frac{2e}{\hbar}\right)^2\int_0^{t}dt_1\int_0^{t}
dt_2\int_0^{t_1}dt'\int_0^{t_2}dt''\langle V(t')V(t'')\rangle.
\end{align}
Since the voltage fluctuations fulfill the Nyquist theorem, we can write
\begin{align}
R=\frac{1}{k_BT}\int_{-\infty}^\infty\langle V(t)V(t+\tau)\rangle d\tau,
\end{align}
where T is the temperature that affects to the external circuit, 
and R is its corresponding resistance. Leading to the result
\begin{align}
\langle\langle\sigma_x\rangle(t)\rangle = \cos(\Delta t)e^{-\frac{1}{2}\gamma_D t^3},
\end{align}
where we have used
\begin{align}
\gamma_D=
\frac{1}{2}\left(\frac{E_J}{\hbar}\right)^2\left(\frac{2e}{\hbar}\right)^2 R k_B T.
\end{align}
Due to the fact that the time needed to vary the phase by 
$\varphi\rightarrow \varphi +2\pi$ is of the order of 
$1/\omega_{ac}$ and $\omega_{ac}=10^{10}$Hz, and taking into 
account that  for the parameters that we have considered 
(see caption of Fig.~3 in the main text) 
$\gamma_{D}/\omega_{ac}^3\gg 1$, coherences become rapidly quenched.

\section{Splitting of the Shapiro steps as a function of $P_{LZ}$}

\subsection{Fourier decomposition of the 4$\pi$-periodic current}

\begin{figure}[tb]
\begin{center}
\includegraphics[width=3.0in,clip]{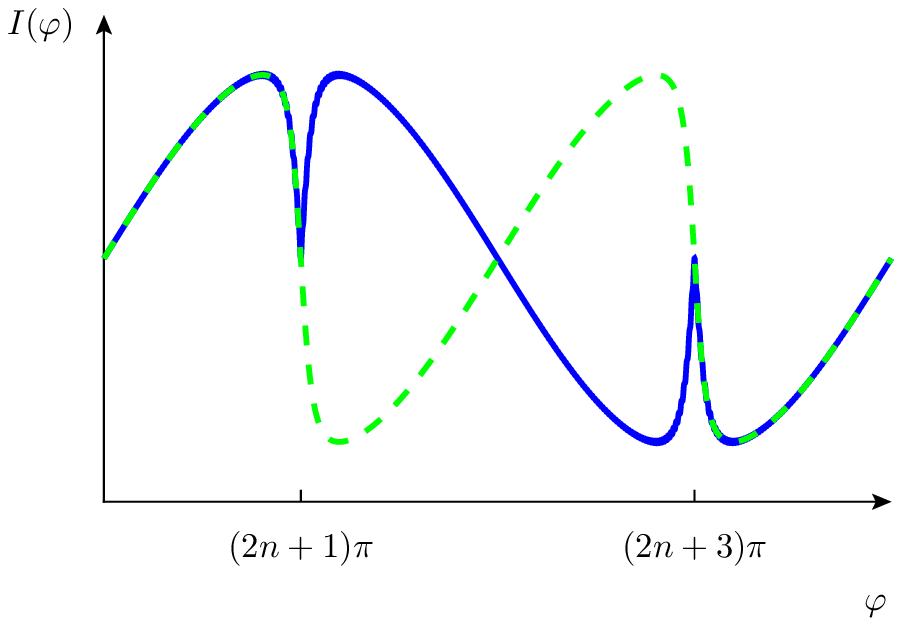}
\end{center}
\caption {\label{Fig.current}\small Dynamical 4$\pi$ 
current (solid) in a Josephson junction after having two 
consequtive LZT. Dahsed curve corresponds to the opposite case, 
i.e.~two consequtive non-LZT.}
\end{figure}

\begin{figure}[tb]
\begin{center}
\includegraphics[width=2.5in,clip]{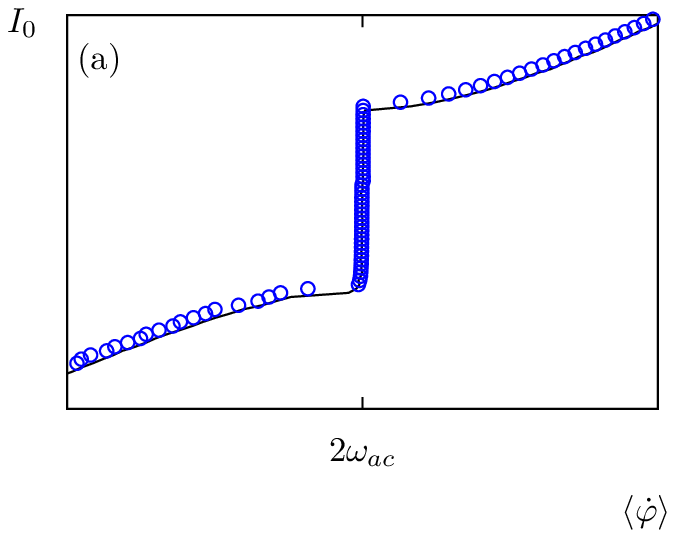}
\includegraphics[width=2.5in,clip]{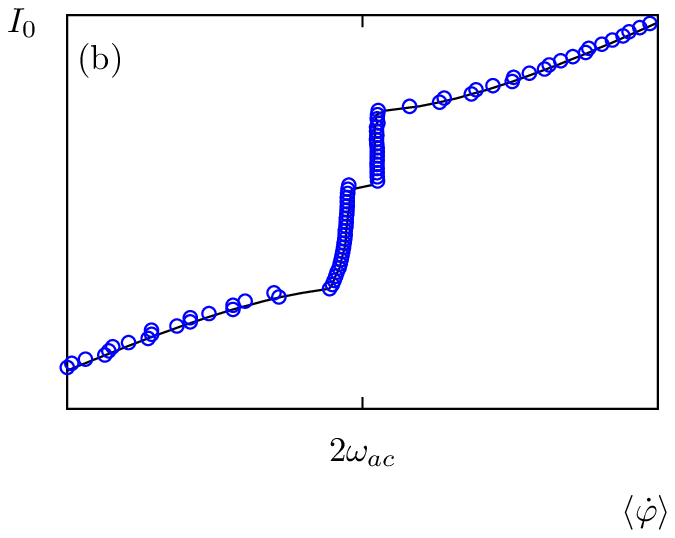}
\includegraphics[width=2.5in,clip]{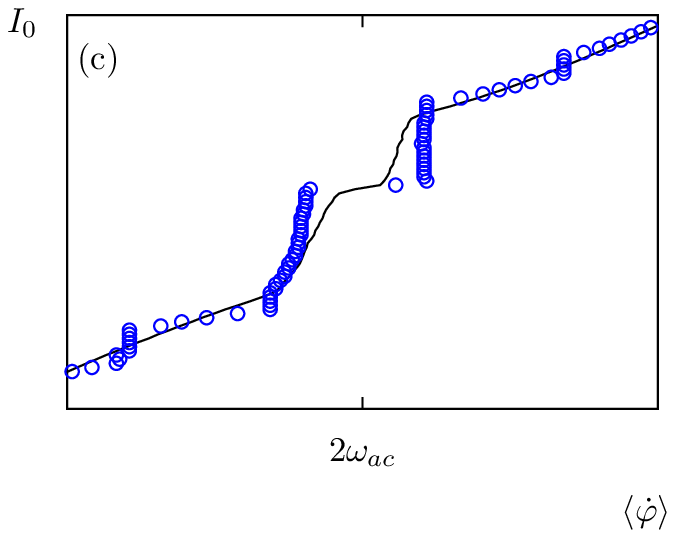}
\includegraphics[width=2.5in,clip]{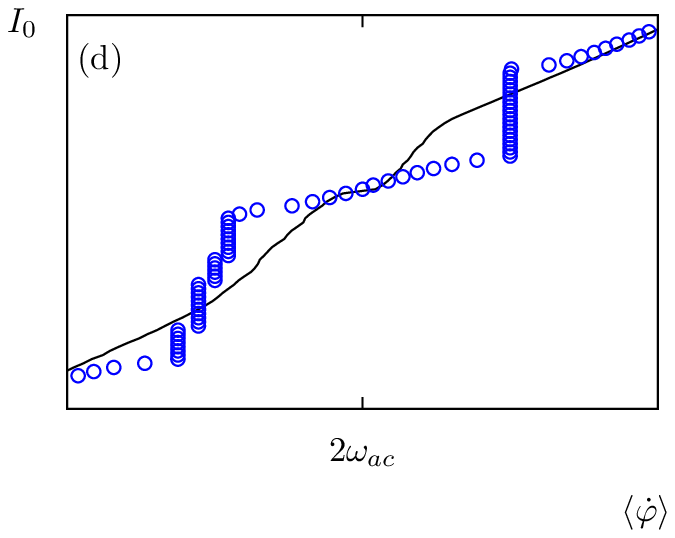}
\end{center}
\caption {\label{Fig.appsplitting}\small Current/voltage curves for different values of
$E_J/\delta$: (a) 500, (b) 100, (c) 50 and
(d) 30, $I_1=1\,$nA, the rest of the parameters
are taken from those used in Fig~3 in the main text. Solid lines are 
the values obtained by means of the stochastic model while blue circles 
come from the averaged model presented here.}
\end{figure}

The current that results of having LZT has the form of the continuous line 
shown in Fig.~\ref{Fig.current}. However, from time to time 
the LZT does not occur, thus, the current suffers changes in its 
periodicity. The consequences of varying the periodicity of the current 
is reflected on the Shapiro steps pattern. It leads to the splitting 
of the steps see Fig~\ref{Fig.appsplitting}. In order to have some analytical 
insight to this phenomena, we use a Fourier expansion of the current and of 
the non-LZT.

We begin calculating the Fourier components of the current when LZT are always
given. We make an odd expansion leading to the terms
\begin{align}
I_F(\varphi)=\sum_{n=0}^\infty b_n \sin\left(\frac{n}{2}\varphi\right) 
\end{align}
where the Fourier components are given by
\begin{align}
b_n=\frac{1}{\pi}\int _0^{2\pi} d\varphi I(\varphi)\text{Sgn}
\left[\cos(\varphi/2)\right] \sin\left(\frac{n}{2}\varphi\right),
\end{align}
here Sgn denotes the sign function and the current 
$I(\varphi)$ is given by

\begin{align}
I(\varphi)=\frac{\partial}{\partial \varphi}E(\varphi)=
\frac{E_J^2\sin(\varphi)}{8\sqrt{4\delta^2 + E_J^2 \cos^2(\varphi/2)}}.
\end{align}
where $\delta$ is the overlap between the in-phase MF and $E_J$ is 
the Josephson energy.
The integral is zero for $n=\dot 2$ yielding just the odd Fourier 
contributions.
Taking into account that $\delta\ll E_J$, we take the first order (Taylor) 
expanssion 
in $\delta$, yielding
\begin{align}
&b_0=\frac{E_J}{4}-\frac{2\delta}{\pi}\\
&b_n=(-1)^{n+1} \frac{2\delta}{\pi}.
\end{align}
This leads to write the current as,
\begin{align}
I_F(\varphi)=\frac{E_J}{4}\sin(\varphi/2)+\frac{2\delta}{\pi} 
\sum_{n=0}^\infty (-1)^{n+1} \sin\left(\frac{2n+1}{2}\varphi\right).
\label{eq.if}
\end{align}
As expected, the current has a 
$4\pi$ contribution proportional to $E_J$ perturbed by a 
sum of extra terms which are of the order of 
$\delta$. 

\subsection{Non-Landau Zener Transitions}

We represent the phenomenon of not having 
LZT in a different way as it is presented in the main text. 
There, having 
a LZT was simulated by a change of sign. Indeed, this has 
been already included in the Fourier expansion above eq.~\ref{eq.if}, 
and for this reason here
having a non-LZT will be simulated by changing 
as well the sign of the current. 
To this aim we will use a unit-step that changes its 
sign after a certain period, which 
depends on the average of the LZ probability.

In the calculations we perform in the main text, 
this effect is 
estimated stochastically. In turn, here we consider 
the average of a 
given LZT and non-LZT. Since a LZT is not affected by a previous 
phase, this approximation will be in principle accurate. 
Thus, we consider the averaged LZ probability $P_{LZ}$, namely, in average
after a number LZT there is a non-LZT. 
The next approximation will be to consider
that these non-LZT are distributed homogenously in time. This approximation,
is a bit more fragile, due to the fact that we are measuring the periodicity 
of the current by the position of the Shapiro steps. 
Thus, to consider a homogeneous distribution or an inhomogeneous
one will lead to different results in the experiment. 
In principle, the homogeneous distribution will be only appropiated
when the non-LZT are separated enough in time, so that the periodicity of 
the 4$\pi$ current (eq.~\ref{eq.if}) dominates.

In addition, the function has to include the fact that the 
change of sign can only come at certain values of the phase, i.e.~when 
$\varphi = (2n+1)\pi$. Then, it is straightforward to obtain the period as
\begin{align}
T=2\pi(2P)
\end{align}
where 
\begin{align}
P=\text{Integer} \left[\frac{1}{1-P_{LZ}}\right]
\end{align}
and therefore the Fourier expansion becomes
\begin{align}
\text{NLZ}(\varphi)=\sum_{i=0}^\infty\frac{4}{(2i+1)\pi}
\sin\left(\frac{2i+1}{2P}(\varphi+\pi)\right)
\end{align}

\subsection{Total current}

Then, taking the product of the Fourier expansions of the current and 
the change of sign and we arrive to the expression 
\begin{align}
I(\varphi)=&\sum_{i=0}^\infty \frac{E_J}{2(2i+1)\pi} 
\left\{\cos\left(\left(\frac{1}{2}-\frac{2i+1}{2P}\right)\varphi\right)-
\cos\left(\left(\frac{1}{2}+\frac{2i+1}{2P}\right)\varphi\right)\right\}\nonumber\\
& +\frac{\delta}{\pi}\sum_{i=0}^\infty \frac{4}{(2i+1)\pi} \sum_{n=0}^\infty 
\left\{ (-1)^{n+1}\left[\cos\left(\left(\frac{2n+1}{2}
-\frac{2i+1}{2P}\right)\varphi\right)-\cos\left(\left(\frac{2n+1}{2}+
\frac{2i+1}{2P}\right)\varphi\right)\right]\right\}.
\end{align}
We can observe that due to the new periodicity the former 4$\pi$ and the 
rest of the Fourier components, are splitted in two by a quantity proportional 
to $2 (1-P_{LZ})$, as occurs in the numerics (see Fig.~\ref{Fig.appsplitting}). 
We can also see that the average model deviates from the stochastic 
calculations as we increase $\delta$.
This is caused because increasing $\delta$, the number of non-LZT increases
so that non-LZT may not occur isoletely, changing thus the periodicity. 
Therefore, for lower LZ probabilities the average between the pure 
2$\pi$ and 4$\pi$ currents weighted by $P_{LZ}$ give more accurate results.

It is worthy to remark that the Fourier current presented above, 
has been developed to show analyticaly the splitting, however, numerical 
results of the averaged model have been obtained from the current given by
\begin{align}
I(\varphi)=I_M\frac{2}{E_J}\frac{\partial}{\partial \varphi}E (\varphi) 
{\text Sgn}\left\{\cos(\varphi/2)\right\}{\text Sgn}
\left\{\cos\left(\frac{\varphi-(P-1) \pi}{2P}\right)\right\}
\end{align}

\section{Shapiro experiment: Robustness of the even steps}

The Shapiro experiment has been proposed 
\cite{Kitaev2001a,Kwon2003a, Lutchyn2010a, Oreg2010a} to 
detect the presence of Majorana fermions
because it allows to deduce the periodicity 
of the current-phase relation of the junction, and therefore
to distinguish between Majorana and normal
modes, whose current is proportional to 
$\sin(\varphi/2)$, and $\sin(\varphi)$ 
respectively.

\begin{figure}[tb]
\begin{center}
\includegraphics[width=2.5in,clip]{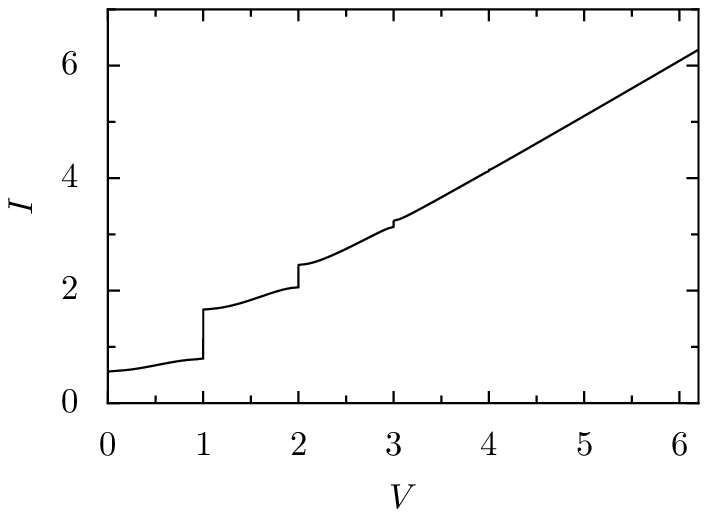}
\includegraphics[width=2.5in,clip]{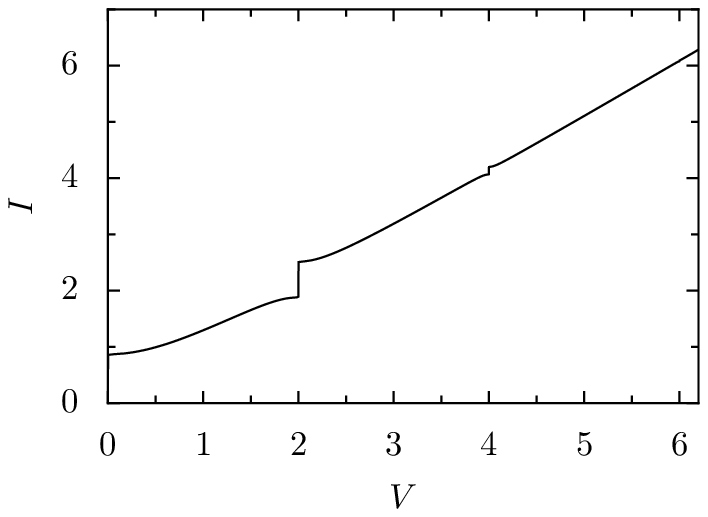}
\end{center}
\caption {\label{Fig.shapiroapp}\small $I/V$ curve of a current biased 
Shapiro experiment in the presence of normal (left)
and Majorana modes (right).}
\end{figure}

The physical fundament of the experiment consists 
of a resonance phenomena which involves 
the natural frequency of the Majorana (normal)
Josephson junction $\omega_0=(2)eV/\hbar$ and 
the frequency of an applied rf current (voltage) $\omega_{ac}$.
When the resonance is fulfilled, 
i.e.~$V=n \hbar\omega_{ac}/(2)e$, Shapiro steps arise. 
We can now see that 
Shapiro steps coming from a Majorana mode arise 
just at even multiples of the applied ac frequency, 
while normal modes appear at 
the whole spectrum of integer multiples, 
see Fig.~\ref{Fig.shapiroapp}.

In practice, to distinguish the presence of a Majorana 
mode is expected to be more complicated 
due to the presence of both, normal and Majorana modes, 
presenting a higher contribution the former one. 
In this scenario, Shapiro steps should appear  
at even and odd positions.
However, in the main text we have found a regime in the 
current biased experiment where the 
Majorana mode can give rise 
to even steps in the presence of a 
much higher 2$\pi$ contribution to the current.
This section is dedicated to explain the reason of this behavior.

The structure of this section consists of two parts. In the first 
one we present briefly the voltage biased experiment, while in the second 
we analyze in more extension the current biased experiment. 
The second part is divided in two subsections where we analyze  
the numerical results of the RSJ model in the absence and presence of 
Majorana and with and without ac current.

\subsection{Voltage biased Shapiro experiment}

In the presence of Majorana and normal modes the Josephson current is given by 
\begin{align}
I(\varphi)=I_M \sin(\varphi/2)+ I_c \sin(\varphi),
\label{eq:Josephcurr}
\end{align}
where $I_M\ll I_c$.
On the other hand, the phase difference is given by the Josephson formula
\begin{align}
\frac{d}{dt}\varphi=\frac{2e}{\hbar}V(t).
\end{align}
This means that applying an external ac-voltage
\begin{align}
V(t)=V_0+V_1\cos(\omega_{ac} t),
\end{align}
we will be able to tune the phase difference, leading to
\begin{align}
\varphi(t)=\varphi_0+\omega_0 t+\frac{2eV_1}{\hbar}\sin(\omega_{ac} t).
\label{eq:phaset}
\end{align}
Here we have used $\omega_0\equiv 2eV_0/\hbar$. 
Then, we substitute eq.~(\ref{eq:phaset}) 
into the Josephson current eq.~(\ref{eq:Josephcurr})
\begin{align}
I(t)=I_M\sum_n (-1)^n J_n\left(\frac{eV_1}{\hbar\omega_{ac}}\right)
\sin\left(\frac{\varphi_0}{2}+\left(\frac{\omega_0}{2}-n\omega_{ac}\right)t\right)
+I_c\sum_n (-1)^n J_n\left(\frac{2eV_1}{\hbar\omega_{ac}}\right)
\sin\left(\varphi_0+(\omega_0-n\omega_{ac})t\right).
\end{align}
Thus, in the stationary limit just the contributions with 
$n\omega_{ac}=\omega_0$ survive, namely
\begin{align}
\bar{ I}=I_M\sum_n (-1)^n J_n\left(\frac{eV_1}{\hbar\omega_{ac}}\right)
\delta\left((\omega_0/2-n\omega_{ac})\right)+I_c\sum_n (-1)^n 
J_n\left(\frac{2eV_1}{\hbar\omega_{ac}}\right)
\delta\left((\omega_0-n\omega_{ac})\right).
\label{eq:linearsteps}
\end{align}
with Dirac deltas placed at integer values of the radio frequency, $n\omega_{ac}$. 
It has to be noted that the contribution of the normal and Majorana modes 
is linear in $I_{c,M}$, and for this reason the height of the even steps 
will be sligthly modified by the presence of the Majorana mode in the case that 
$I_c \gg I_M$.
Therefore, we can understand that the presence of the normal Andreev modes, which
contributes with steps placed at all integer multiples, will difficult the 
separation of both contributions and thus the identification of the Majorana mode. 
For this reason, in the situation where $I_c \gg I_M$, the voltage biased 
experiment seems to be a non-sensitive method to detect the Majorana mode.

\subsection{Current biased Shapiro experiment}

In the main text we have performed numerical calculations for the current 
biased Shapiro experiment. We have presented I-V curves for 
different values of $\omega_{ac}$ and seen that for some special regime
of frequencies odd steps vanish, pointing out the 4$\pi$ periodicity of 
the junction. 
In this section we give numerical arguments to explain this phenomena.
In order to simplify the discussion we will study the ideal case, 
where the Majorana fermions at the extremes of the quantum wire 
are infinitely apart
from the junction, so that the overlap is zero and the supercurrent 
becomes proportional
to $\sin(\varphi/2)$. In this situation the equation under study is 
\begin{align}
I_0+I_1\sin(\omega_{ac} t)=I(\varphi(t))+\frac{\hbar}{2e R}\dot{\varphi}(t).
\label{eq:appRSJ}
\end{align}
This equation is obtained from Kirchoff's law where an external dc $I_0$ and
ac $I_1 \sin(\omega_{ac} t)$ currents are applied to the junction. The
outgoing current is modeled by a parallel circuit whose components are,
$I(\varphi(t))$, given by Eq.~(\ref{eq:Josephcurr}), and a resistive current
$\hbar/(2e R)\dot{\varphi}$ originating from the existence of quasiparticles.

In order to analyze the equation it results convenient to renormalize the
involved parameters. 
Therefore, we divide the entire equation by $I_c$, and transform the time to 
a dimensionless
quantity $\tau=(2e R/\hbar )t$. This provokes a renormalization on the 
ac frequency $\xi=\hbar\omega_{ac}/2eR I_c$, leading to rewrite the equation as
\begin{align}
\dot\varphi(\tau)=\alpha_0+\alpha_1\sin(\xi \tau)-\sin(\varphi)-\alpha_M \sin(\varphi/2)
\label{eq:rsjrenorm}
\end{align}
where the renormalized intensities are given by $\alpha_i=I_i/I_c$. 
Now, the analysis of the equation is reduced to the study the parameter 
regime of $\alpha_1,~\alpha_M$, and $\xi $. 

The analytic solution of the equation in the absence of the Majorana mode 
is not known, and only approximate solutions have been 
obtained \cite{Aslamazov1969a, Thompson1973a, Kvale1990a}, although none of
them are valid in the regime where we observe that odd steps vanish.
Besides, the 4$\pi$ term coming from 
the presence of the Majorana mode 
makes the system even more complicated.
For all these reasons, it is out of the scope 
of this appendix to try to give an analytical insigth 
of the differential equations. In turn, we 
explore the numerical solutions and 
explain its general behavior.

\subsubsection{Undriven system without the Majorana term}

\begin{figure}[tb]
\begin{center}
\includegraphics[width=3.0in,clip]{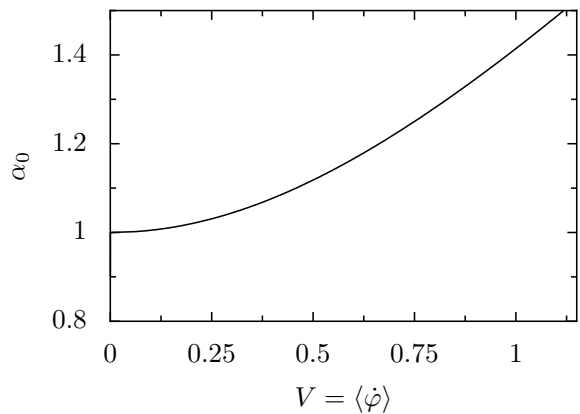}
\end{center}
\caption {\label{Fig.a0V}\small $\alpha_0$ vs $\langle \dot\varphi\rangle$ 
in the absence of the Majorana mode and the ac current.}
\end{figure}

The equation of motion of the system without neither MF nor 
ac current is given by
\begin{align}
\dot\varphi(t)=\alpha_0-\sin(\varphi(t)).
\end{align}
In Fig.~\ref{Fig.a0V} we show the dependence of 
$\langle\dot\varphi\rangle$ as a function of $\alpha_0$.
This case is analytically solvable, and the average of the voltage
is given by
\begin{align}
\bar{V}=\langle \dot\varphi \rangle=\sqrt{\alpha_0-1}.
\end{align}
This equation is only valid for $\alpha_0>1$. The value
$\alpha_0=1$ is called the critical value, which we denote by 
$\alpha_c$, and is 
defined as the value up to which the induced voltage passes 
from zero to a finite value.
In general, this value will change depending 
on whether we add the Majorana mode and/or the ac current.
This value will be important in the incoming analysis.

In panel a of Figs.~\ref{Fig.4pi} and~\ref{Fig.4pia0} 
we show numerical results of $\varphi(\tau)$ and $\dot\varphi(\tau)$ 
as a function of time for two different values of $\alpha_0$. 
Figure~\ref{Fig.4pi} presents values of $\alpha_0$ close to 
the critical value of each system, while Fig.~\ref{Fig.4pia0}
presents a higher value of $\alpha_0$.
As we can observe, $\varphi(\tau)$ presents 
two different ranges of slopes, i.e. velocities, fast and slow. 
The slow range is given when the term $\sin(\varphi)= 1$,
so that the difference $\alpha_0-\sin(\varphi)$ is minimum. 
This happens when $\varphi=(4n+1)\pi/2$ (dashed lines of panels a and b of  
Figs.~\ref{Fig.4pi} and~\ref{Fig.4pia0}). 
In figure~\ref{Fig.4pi} $\alpha_0$ is very close to $\alpha_c$,
thus the difference $\alpha_0-\sin(\varphi)\rightarrow 0$, 
and therefore we find a flat slope. For larger values 
of $\alpha_0$ (see Fig.~\ref{Fig.4pia0}) the difference does not 
tend to zero and we observe an
increment of the slope between stairs. 
In the bottom plot of Figs.~\ref{Fig.4pi} 
and~\ref{Fig.4pia0} we show 
results of $\dot \varphi$ vs. time. 
There we can observe that the increment 
of $\alpha_0$ induces a widening of the peaks.
On the other hand, 
the fast range is given for the rest of the values of $\varphi(\tau)$, 
having a maximum speed when $\sin(\varphi(\tau))=-1$ which occurs at
$\varphi=(4n+3)\pi/2$.
These ranges are periodically repeated with the frequency
\begin{align}
\omega_0=\bar{V}.
\end{align}
in units of $2e/\hbar$.
That is, the frequency of the junction is proportional to the 
induced voltage. 
In summary, by increasing $\alpha_0$ we modify the 
frequency of the junction $\omega_0$, and we also provoke
an increment of the slope between stairs, which produces 
a widening of the peaks seen in $\dot\varphi(\tau)$.

\subsubsection{Undriven system with the Majorana mode}

\begin{figure}[tb]
\begin{center}
\includegraphics[width=2.0in,clip]{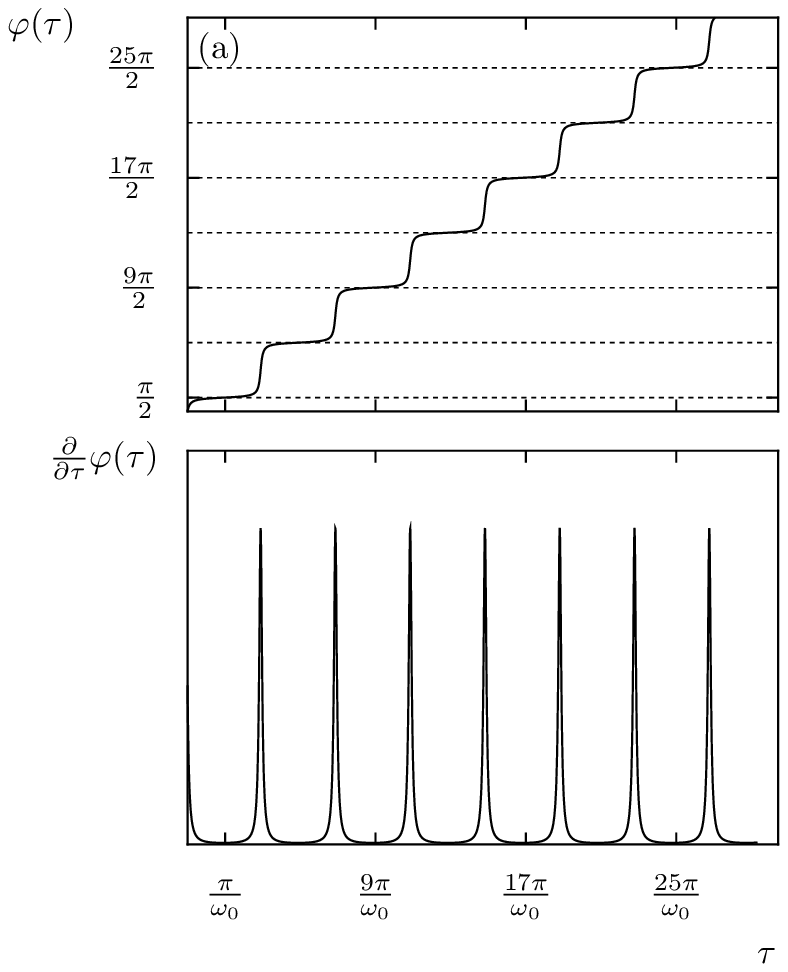}
\includegraphics[width=2.0in,clip]{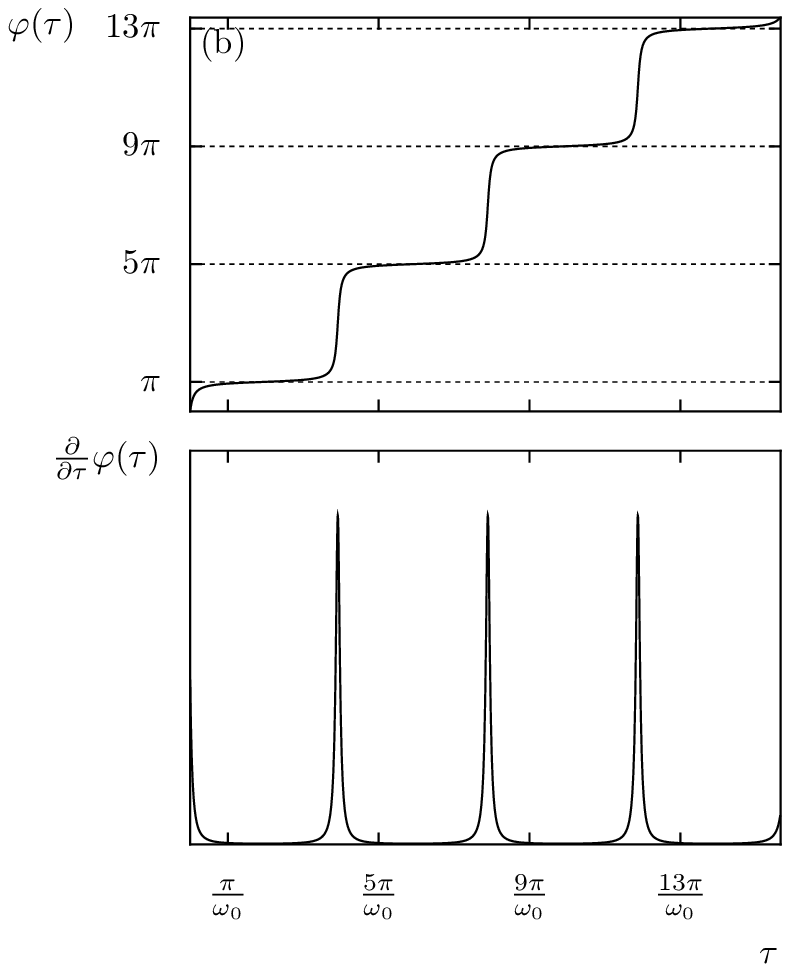}
\includegraphics[width=2.0in,clip]{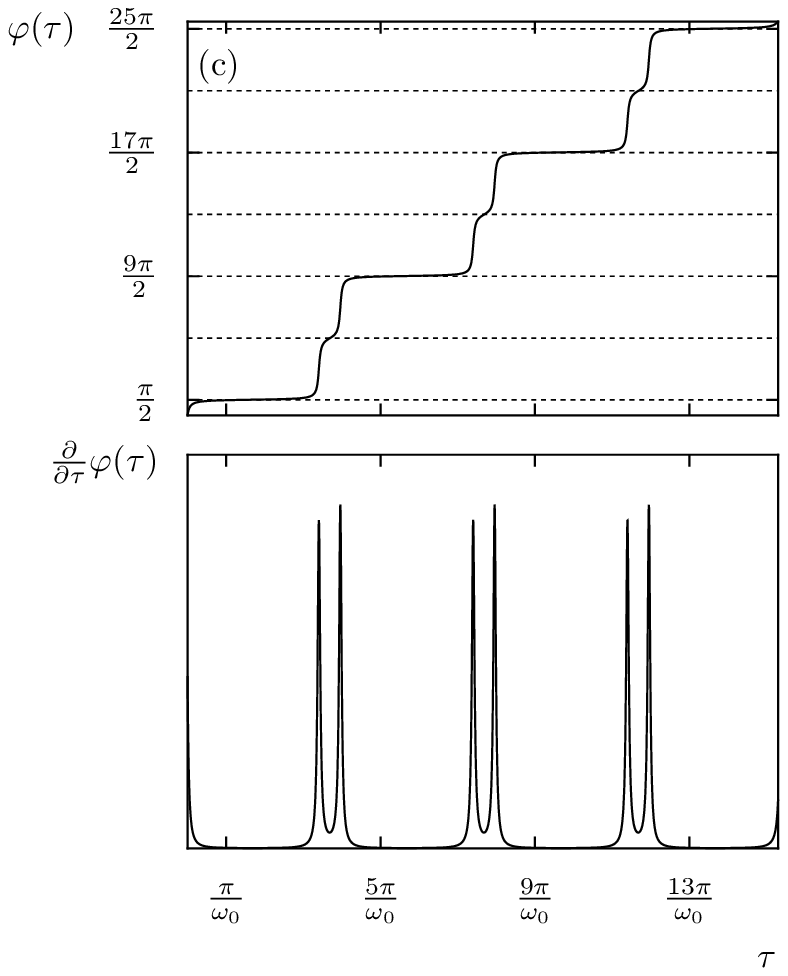}
\end{center}
\caption {\label{Fig.4pi}\small $\varphi(\tau)$ (top row) and $\dot\varphi (\tau)$ 
(bottom row)
as a function of 
time with a value of $\alpha_0$ close to the critical value 
$\alpha_c$ in a 2$\pi$, 4$\pi$ and mixed 
situation with $\alpha_M= 1/15$, from left to right.}
\end{figure}

\begin{figure}[tb]
\begin{center}
\includegraphics[width=2.0in,clip]{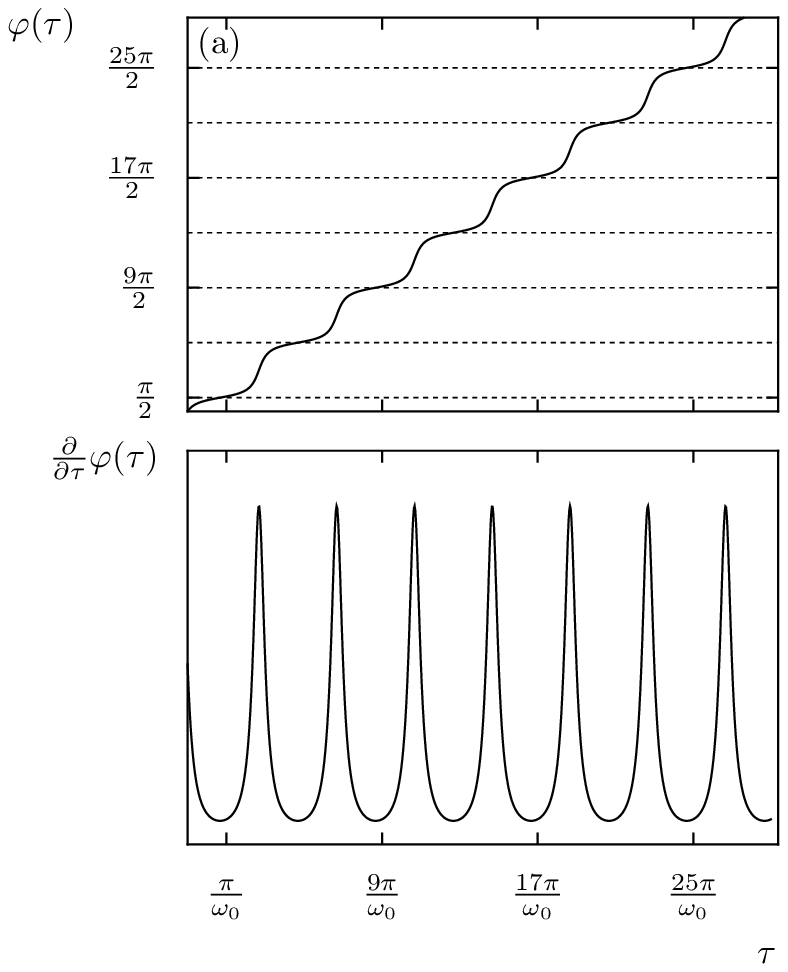}
\includegraphics[width=2.0in,clip]{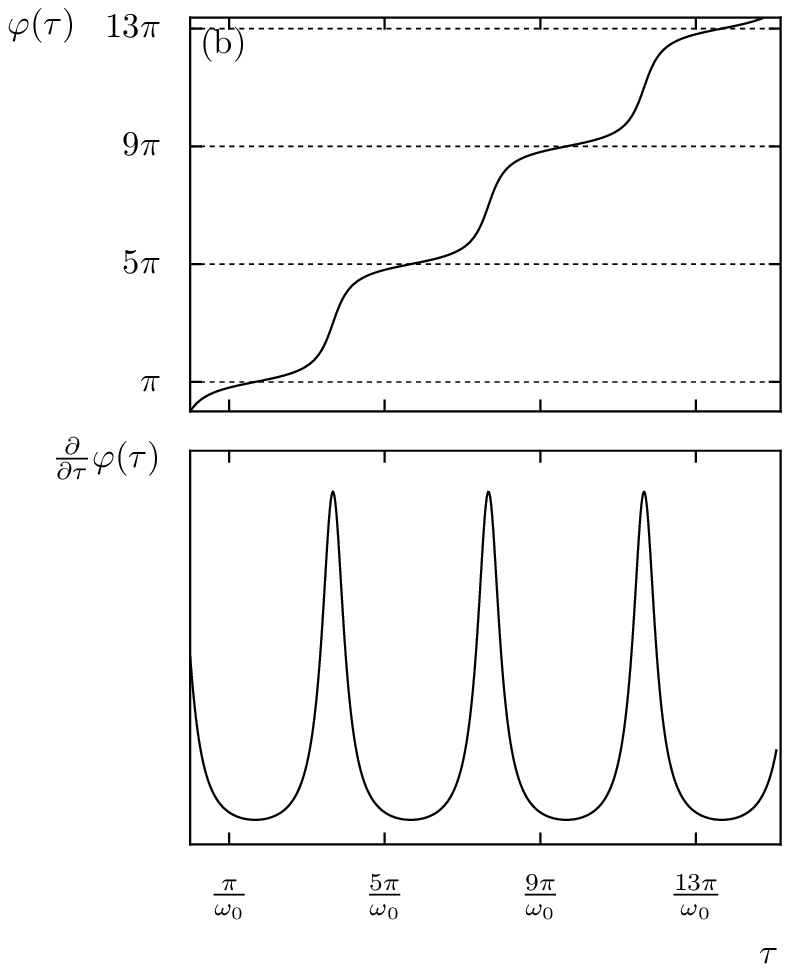}
\includegraphics[width=2.0in,clip]{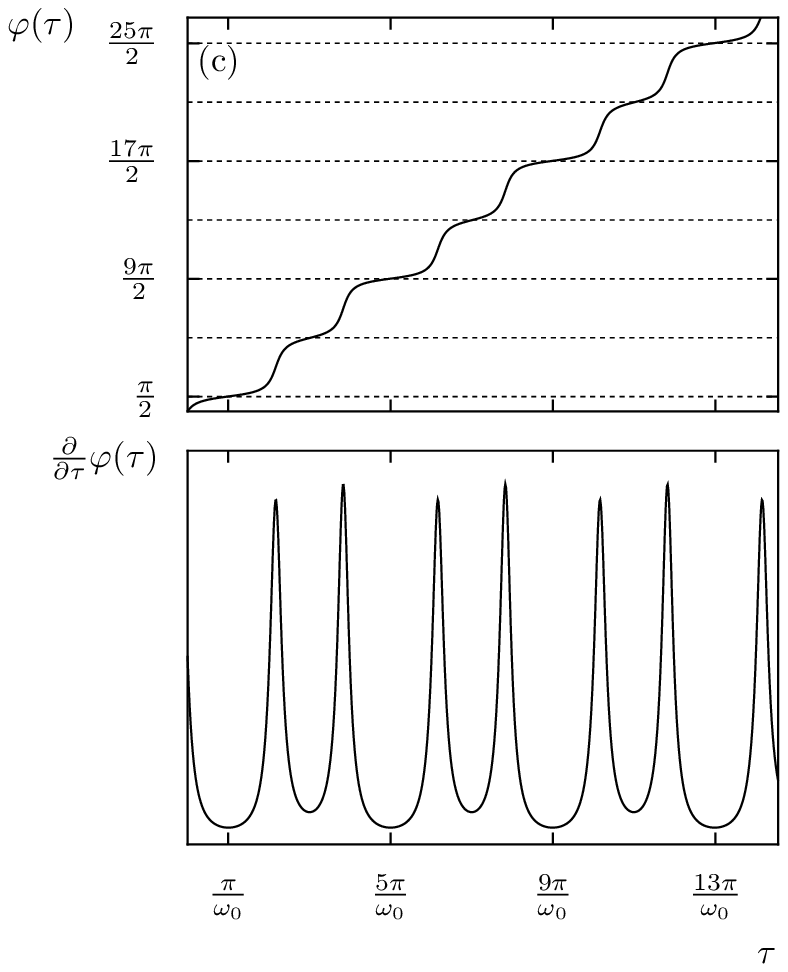}
\end{center}
\caption {\label{Fig.4pia0}\small Same plot as Fig.~\ref{Fig.4pi} 
for a larger value of $\alpha_0$ respect to the critical value $\alpha_c$.
Comparing panel c of this figure and Fig.~\ref{Fig.4pi} we can appreciate the 
change of 
period when we move away from $\alpha_c$, i.e.~ when $\alpha_0\rightarrow \alpha_c$ 
(Fig.~\ref{Fig.4pia0}c$\rightarrow$Fig.~\ref{Fig.4pi}c) the periodicity of 
$\dot\varphi(\tau)$ approaches 4$\pi$}
\end{figure}

We add now the Majorana mode by including the term 
$\alpha_M\sin(\varphi /2)$ to the supercurrent yielding
\begin{align}
\dot\varphi(\tau)=\alpha_0-\sin(\varphi)-\alpha_M\sin(\varphi/2),
\end{align}
with $\alpha_M \ll 1$.
In the example presented here we have used 
the value $\alpha_M=1/15$, so that
the sum of this term modifies slightly 
$\sin(\varphi)$. However, as we can observe in panel c
of Fig.~\ref{Fig.4pia0}, the periodicity of 
$\varphi(\tau)$ is drastically modified for values of 
$\alpha_0$ close to the new critical value, 
$\alpha_c\approx 1+\alpha_M/\sqrt{2}$ (see panel c of 
Fig.~\ref{Fig.4pi}). The solution $\varphi(\tau)$ turns from 
a $2\pi$ periodicity to 
a 4$\pi$ for values of $\alpha_0$ close to the critical value $\alpha_c$ 
(see Fig.~\ref{Fig.4pi}c) and 
becomes 2$\pi$ for larger values of $\alpha_0$ (see Fig.~\ref{Fig.4pia0}c). 
This effect is non-linear and makes a
difference respect to the voltage 
biased experiment.

The change in the periodicity can be explained
by means of analogous arguments as above. 
The addition of the Majorana mode may 
increase or decrease the duration of 
the flat regions depending on the sign of $\sin(\varphi/2)>0$. 
The time difference between the 
short and long periods increases as long as we are closer to 
the critical value $\alpha_c$.
Thus, the 2$\pi$ periodic 
function turns to a 4$\pi$ periodic one as $\alpha_0\approx \alpha_c$, 
still with the characteristic frequency $\omega_0$. 
A more visual comparison can be made just by looking the 
similarities with a pure $4\pi$ Josephson junction
in panel b of Figs.~\ref{Fig.4pi} and~\ref{Fig.4pia0}.

\begin{figure}[tb]
\begin{center}
\includegraphics[width=3.0in,clip]{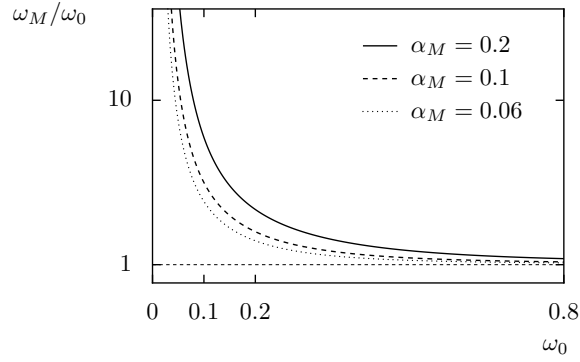}
\end{center}
\caption {\label{Fig.freqs4pi}\small Relation $\omega_M/\omega_0$ vs. $\omega_0$
for $\alpha_M=1/15,~1/10$ and 1/5 (from left to right). 
It represents the 4$\pi$-contribution of the junction as a function of the 2$\pi$, 
given by $\omega_0$. The range of values where  
$\omega_M/\omega_0\gg 1$ indicates the 4$\pi$ behavior of the junction. 
The dashed line corresponds to the value 
$\omega_M/\omega_0= 1$, where is pure 2$\pi$ periodic.}
\end{figure}

We can study continously the transition from 
4$\pi$ to the 2$\pi$ character of the junction calculating from numerics
the largest frequency $\omega_M$ of the junction, defined by 
$\omega_M=2\pi/\Delta \tau$, where $\Delta \tau$ is the dimensionless 
time difference between the two closest maximums of $\dot \varphi(\tau) $.
When $\omega_M$ approaches $\omega_0$, the junction turns to be 2$\pi$ periodic. The
general behavior is shown in Fig.~\ref{Fig.freqs4pi}, where we have plotted the relation 
$\omega_M/\omega_0$ vs. $\omega_0$ for 
three different values of $\alpha_M=1/15,~1/10$ and 1/5. 
The curves show the tendency of the 
periodicity of the junction as 
we increase $\omega_0$. We can observe that 
there is always a range of values of 
$\alpha_0$, i.e.~a range of $\omega_0$, 
close to $\alpha_c$ where the Majorana mode 
imposes its 4$\pi$ periodicity to the junction. Roughly
speaking, this range is of the order of $\alpha_M$. The dashed 
curve placed at $\omega_M/\omega_0=1$ points out the tendency of the junction 
to behave 2$\pi$ periodically.


\subsubsection{Driven system with the Majorana mode}

Until now we have seen that the phase presents a range of voltages or frequencies 
where its dynamics is governed by the periodicity of the Majorana mode. The 
question now is that if we will be able to measure the periodicity of the phase 
in that range of frequencies. Typical procedure to measure the periodicity implies 
to drive the phase by means of an ac current, that is
\begin{align}
\dot\varphi(\tau)=\alpha_0+\alpha_1\sin(\xi \tau)-\sin(\varphi(\tau))-
\alpha_M\sin(\varphi/2),
\end{align}
It is well known that the 
solutions $\varphi(\tau)$ at the steps are 
phase-locked solutions (e.g~see Ref.~\onlinecite{Thompson1973a}). 
This means that the driving force imposes
its frequency to the driven system. 
And in this way, the solutions of $\dot \varphi(\tau)$ change from 
the former frequency imposed by $\alpha_0$, i.e.~$\omega_0$, 
to the ac frequency $\xi$. 

\begin{figure}[tb]
\begin{center}
\includegraphics[width=3.0in,clip]{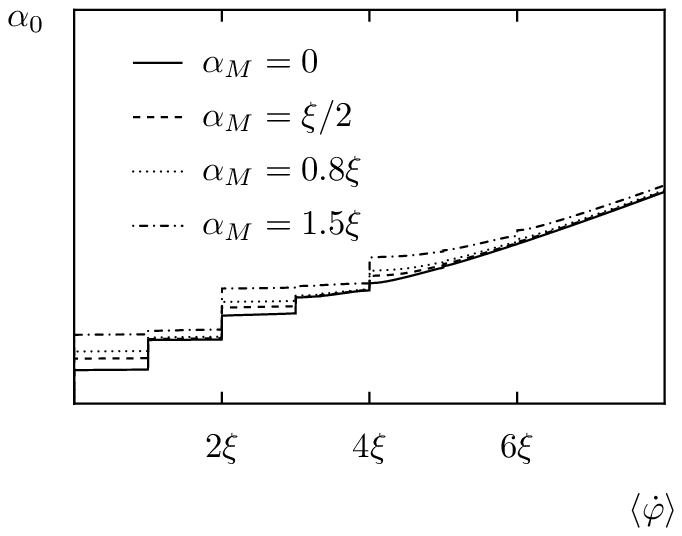}
\includegraphics[width=3.0in,clip]{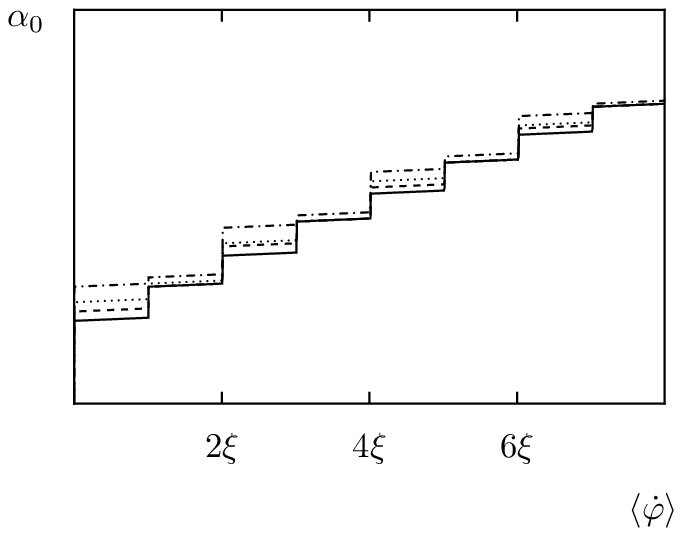}
\end{center}
\caption {\label{Fig.Shapiro-am}\small Shapiro I-V curves for $\alpha_1=0.3$, 
and $\xi=1/5$ for the left plot and $\xi=1/20$ for the right plot. 
The value of $\alpha_M$ is increased from 0 to $1.5\xi$ in both plots. 
We can appreciate the reduction of the odd steps as long as 
$\alpha_M \gtrsim \xi$.}
\end{figure}

\begin{figure}[tb]
\begin{center}
\includegraphics[width=3.0in,clip]{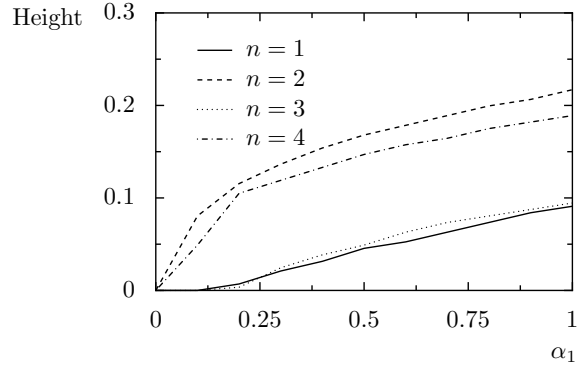}
\end{center}
\caption {\label{Fig.alturas}\small Height of the first four 
steps as a function of $\alpha_1$ for 
$\alpha_M=0.15$ and $\xi=0.1$.}
\end{figure}

This special property of the phase locked solutions 
is very important because it
allows to access the range of frequencies where 
the junction behaves 4$\pi$ periodically: 
Choosing $\alpha_M\gtrsim\xi$, that is, the order where we have 
seen that $\omega_M/\omega_0\gg 1$, 
will lead to have a dominant 4$\pi$ periodicity, 
which leads to an even Shapiro spectrum. 
We can see this behavior in the plots shown 
in Fig.\ref{Fig.Shapiro-am}, where we have plotted 
the Shapiro steps for different values of $\alpha_M$. 
We see how the 
odd steps tend to vanish for $\alpha_M\approx\xi$.
Finally, we have plotted in Fig.~\ref{Fig.alturas} 
the height of the first four steps 
as a function of the ac intensity $\alpha_1$ for $\alpha_M =0.15$ and $\xi=0.1$. 
We can observe a clear 
predominance of the even steps, for the whole range of $\alpha_1$. 
Remarkably, we see that for $\alpha_1\approx\alpha_M$, odd steps are zero. 
This behavior can be explained by the fact that in our reasoning we have implicitely 
considered that $\alpha_1$, reads out the periodicity of the junction at the imposed
frequencies $\xi$. In other words, we have considered that the effect of adding the 
ac current is to select the frequency of the junction, without introducing its 
2$\pi$ periodicity, and in such sense $\alpha_1$ needs to be of the order of $\alpha_M$.
We see that for larger values of $\alpha_1$, odd steps coming from a 
2$\pi$ contribution become larger.

\subsubsection{Conclusions}

In this section we have explained numerically 
the behavior of the solutions $\varphi(\tau)$ and $\dot\varphi(\tau)$
of the current biased Shapiro experiment in the presence of a
Majorana and several normal modes. 
We have given the explanation of the dominance 
(for a range of frequencies) of the 
even Shapiro steps in the presence of an, in principle, 
negligible 4$\pi$ contribution.
In order to understand this behavior we have explored 
the solutions of the undriven 
system and seen that in general there is always a region 
of frequencies where the 4$\pi$ periodicity dominates. 
This region has an interval of frequencies of the order of 
$\omega_0\lesssim\alpha_M$, 
where $\alpha_M$ is the dimensionless intensity of the Majorana mode.
Therefore, if one wants to measure some signature of the 
4$\pi$ periodicity of the system,
it will be needed that the measurement is performed in this frequency 
regime.

One of the advantages of the current biased Shapiro experiment consists 
on the fact that the steps present phase locked solutions. This means
that the forced system imposes its periodicity to the junction. 
Therefore,
we can impose the ac frequency $\xi$ to the junction $\omega_0$
by means of biasing the junction by an ac current. And meanwhile
measure the periodicity of the junction by looking at the positions
of the steps.
We have seen that when $\alpha_M\gtrsim\xi$
even Shapiro steps dominate, and also that this behavior is more robust when 
$\alpha_1\approx \alpha_M$. Transforming back to physical units 
we have that taking into account that $I_M=1\,$nA and $R=3\,$k$\Omega$, then 
$\omega_{ac}\lesssim 2eR I_M/\hbar =10^{10}\,$Hz.



\end{document}